\documentstyle[pra,aps]{revtex}
\begin{document}
%\draft

\title{\bf Entanglement and the $SU(2)$ phase states in atomic
systems}

\author{M. Ali Can, Alexander A. Klyachko, and Alexander S. Shumovsky}

\address{Faculty of Science, Bilkent University, Bilkent, Ankara,
06533, Turkey \\ quant-ph/0202041, February, 2002}

\maketitle

\begin{abstract}
We show that a system of $2n$ identical two-level atoms
interacting with $n$ cavity photons manifests  entanglement and
that the set of entangled states coincides with the so-called
$SU(2)$ phase states. In particular, violation of classical
realism in terms of the GHZ and GHSH conditions is proved. We
discuss a new property of entanglement expressed in terms of local
measurements. We also show that generation of entangled states in
the atom-photon systems under consideration strongly depends on
the choice of initial conditions and that the parasitic influence
of cavity detuning can be compensated through the use of Kerr
medium.
\end{abstract}

\pacs{PACS number(s): 03.65.Ud, 42.50.Ct, 03.67.-a}

\narrowtext

\twocolumn

\newpage

\section{Introduction}

It has been recognized that entanglement phenomenon touches on the
conceptual problems of reality and locality in quantum physics as
well as the more technological aspects of quantum communications,
cryptography, and computing. In particular, the methods of quantum
key distribution in communication channels secured from
eavesdropping are based on the use of entangled states
\cite{1,2,3,4,5} (for recent review, see Refs. \cite{6,7}). In
turn, the realization of quantum computer \cite{8} is dependent on
the ability to form entangled states of initially uncorrelated
single-particle states \cite{9}.

In recent years, many successful experiments have been performed
to verify the violation of Bell's inequalities and
Greenberger-Horne-Zeilinger (GHZ) equality \cite{10,11} and to
develop the methods of engineered entanglement for quantum
cryptography and quantum key distribution. In particular, the
recent advances in the field of cavity QED and techniques of atom
manipulation, trapping, and cooling enable a number of experiments
which investigates the entanglement in the atomic systems (see
\cite{11,12,13,14,15,16,17} and references therein).

It has been shown recently \cite{18} that a pure entangled state
of two atoms can be obtained in an optical resonator through the
exchange by a single photon. The main idea in Ref. \cite{18} is
that a single excitation of the system is  either carried by a
photon or shared between the atoms. If the photon can leak out
from the resonator, the absence of photon counts in the process of
continuous monitoring of the cavity decay can be associated with
the presence of the pure entangled atomic state. The importance of
this scheme is caused by the fact that its realization seems to be
easy available with present experimental technique.

The main objective of this paper is to show that the entangled
states in the "atoms plus photons" systems of the type discussed
in \cite{18} can be represented by the so-called $SU(2)$ phase
states corresponding to the $SU(2)$ algebra of the odd "spin"
\begin{eqnarray}
j= \frac{1}{2} \left[ \left( \begin{array}{c} 2n \\ n \end{array}
\right) -1 \right] ,  \label{1}
\end{eqnarray}
where $2n$ is the even number of atoms and $n=1,2, \cdots$ is the
number of cavity photons. In particular, the system considered in
Ref. \cite{18} corresponds to the phase states of "spin" $j=1/2$.
The $SU(2)$ phase states were introduced in \cite{19} for an
arbitrary spin and then generalized in \cite{20,21} to the case of
the $SU(2)$ subalgebra in the Weyl-Heisenberg algebra of photon
operators (for recent review, see \cite{22}). From the
mathematical point of view, this is the system of
\begin{eqnarray}
 N=2j+1 \nonumber
\end{eqnarray}
qubits defined in the Hilbert space
\begin{eqnarray}
{\cal H}_N=({\bf C}^2)^{\otimes N} \nonumber
\end{eqnarray}
with the componentwise action of $SU(2)^N$. In particular, we show
that these states violate the classical realism and discuss their
realization.

On the other hand, we will discuss a new condition of entanglement
has been proposed recently \cite{23}. Let us note in this
connection that, in the usual treatment of entanglement, the
entangled states of a two-component (in general, multi-component)
system are considered as the nonseparable states with respect to
the subsystems (e.g., see \cite{24}). For example, if the
individual components of a two-component system are described by
the states $|\xi_i \rangle$ and $|\chi_i \rangle$, respectively,
the state
\begin{eqnarray}
| \psi_{ent} \rangle = \sum_i b_i | \xi_i \rangle \otimes | \chi_i
\rangle , \nonumber \\ \langle \xi_i | \xi_k \rangle = \langle
\chi_i | \chi_k \rangle = \delta_{ik}, \quad \sum_i |b_i|^2=1
\nonumber
\end{eqnarray}
is entangled if $b_i \neq 0$ for at least two distinct values of
the subscript $i$. From the mathematical point of view, the
entanglement is caused by the combination of the superposition
principle in quantum mechanics with the tensor product structure
of the space of state of the two- or multi-component system
\cite{25}.

Very often, the existence of entanglement is verified in terms of
violation of Bell's inequalities and their generalizations
\cite{26,27,28,29,30,31}. Another way is based on the use of GHZ
theorem \cite{10}. A possibility to introduce more general
inequalities is also discussed \cite{32}.

It should be noted that the use of Bell's inequalities and their
numerous generalizations demonstrate nothing but the nonexistence
of hidden variables. Moreover, it is possible to say that the
unique, general, and mathematically correct definition of
entanglement still does not exist (e.g., see Ref. \cite{32}).

An interesting approach has been proposed recently \cite{32}.
Considering the state shared between Alice and Bob as a quantum
communication channel, the authors of Ref. \cite{32} concluded
that the information in the case of entanglement is carried mostly
by the correlations between the ends of the channel. These
correlations manifest themselves by means of the local
measurements on the sides of the channel \cite{23}.

Following \cite{23}, consider a composite system defined in the
Hilbert space
\begin{eqnarray}
{\cal H}= \bigotimes_{\ell} {\cal H}^{(\ell)}, \quad \ell \geq 2.
\nonumber
\end{eqnarray}
Let $G$ be the group of dynamical symmetry of a subsystem in the
composite system. Then the Hermitian operators $g$ associated with
representation of $G$ in ${\cal H}^{(\ell)}$ define the set of
local measurement on the corresponding side of the channel
provided by a state $| \psi \rangle \in {\cal H}$. For example, in
the case of ${\cal H}^{(\ell)}={\bf C}^2$, corresponding to the
EPR spin-$\frac{1}{2}$ system, $G=SU(2)$ and the set of local
measurements can be specified by the infinitesimal generators of
the $SL(2)$ group
\begin{eqnarray}
\{ g \} = \{ \sigma^{(\ell)}_k \} , \quad k=1,2,3 , \nonumber
\end{eqnarray}
which is the complexification of the $SU(2)$ group.

It was shown in \cite{23} that the maximum correlation between the
ends of the channel corresponds to the states such that
\begin{eqnarray}
\forall g , \quad \langle g \rangle =0. \nonumber
\end{eqnarray}
This statement can be illustrated by the atoms-plus-photons
systems under consideration. Consider first the set of two
identical two-level atoms. Let $|e_{\ell} \rangle$ and $|g_{\ell}
\rangle$ denote the excited and ground atomic states of the
$\ell^{th}$ atom, respectively. Then, the entangled, maximum
excited atomic states in the system "$2$ atoms plus $1$ photon"
considered in \cite{18} are
\begin{eqnarray}
| \psi_{\pm} \rangle = \frac{1}{\sqrt{2}} (|e_1g_2 \rangle \pm
|g_1e_2 \rangle ) . \label{2}
\end{eqnarray}
Then, the local measurement $g$ can be described by the Pauli
matrices
\begin{eqnarray}
\sigma_1^{(\ell)} & = & |e_{\ell} \rangle \langle g_{\ell}
|+|g_{\ell} \rangle \langle e_{\ell} |, \nonumber \\
\sigma_2^{(\ell)} & = & -i|e_{\ell} \rangle \langle g_{\ell}
|+i|g_{\ell} \rangle \langle e_{\ell} |, \nonumber \\
\sigma_3^{(\ell)} & = & |e_{\ell} \rangle\langle
e_{\ell}|-|g_{\ell} \rangle \langle g_{\ell}| , \label{3}
\end{eqnarray}
i.e., by the infinitesimal generators of the algebra $SL(2)$.  It
is now a straightforward matter to check that
\begin{eqnarray}
\forall i, \ell \quad \langle \psi_{\pm}| \sigma_i^{( \ell )}|
\psi_{\pm} \rangle =0 , \label{4}
\end{eqnarray}
where averaging is taken over the states (4). Another example is
provided by the GHZ states \cite{10}
\begin{eqnarray}
| \psi^{(GHZ)}_{\pm} \rangle = \frac{1}{\sqrt{2}} (|e_1e_2e_3
\rangle \pm |g_1g_2g_3 \rangle ), \label{5}
\end{eqnarray}
corresponding to the maximum atomic excitation in the
$3+3$-system. It is easily seen that the averaging of the local
operators (3) over (5) gives the same result as (4).

This property (4) can be used to define the entangled states.

We will show that the $SU(2)$ phase states of spin $j$ defined by
Eq. (1) in a $2n+n$-type atom-photon system obey the
nonseparability conditions, have the property
(4), and manifest the violation of classical realism expressed in
terms of the GHZ \cite{10} and CHSH (Clauser-Horne-Shimoni-Holt)
\cite{33} conditions.

The paper is organized as follows. In Sec. II, we consider the
representation of the $SU(2)$ phase states. As a particular
example, we examine the system of two identical two-level atoms,
interacting with a single cavity photon and show that the maximum
entangled atomic states of the Ref. \cite{18} belong to the class
of the $SU(2)$ phase states of spin $j=1/2$. Let us stress that
hereafter the maximum entanglement is defined  in the usual way by
the maximum of reduced entropy (e.g., see Refs.
\cite{23,25,27,32}). Then, we generalize this result on the case
of $2n+n$ system. As a nontrivial example we consider in Sec. III
the system of four identical two-level atoms interacting with the
two cavity photons. In this case, the set of entangled, maximum
excited atomic states is provided by the six orthogonal $SU(2)$
phase states of spin $j=5/2$. For these states, we prove violation
of classical realism through the use of GHZ and CHSH conditions.
In Sec. IV, we discuss how the entangled atomic states can be
achieved in the process of steady-state evolution. In particular,
we show that the maximum entanglement can be achieved if the
initial state of the system contains the photons and does not
contain the atomic excitations. We also show that the presence of
the cavity detuning hampers the creation of pure entangled states
and that the parasitic influence of detuning can be compensated
through the use of the Kerr medium inside the cavity. Finally, in
Sec. V, we briefly discuss the obtained results.

\section{Representation of the $SU(2)$ phase states}

An arbitrary spin $j$ can be described by the generators
$J_+,J_-,J_z$ of the $SU(2)$ algebra such that
\begin{eqnarray}
[J_+,J_-]=2J_z, \quad \quad [J_z,J_{\pm}]= \pm J_{\pm}, \nonumber
\\ J^2=J_z^2+ \frac{1}{2} (J_+J_-+J_-J_+)=j(j+1) \times {\bf 1},
\label{6}
\end{eqnarray}
where ${\bf 1}$ is the unit operator in the $2j+1$ dimensional
Hilbert space. Since
\begin{eqnarray}
J_{\pm}= J_x \pm iJ_y, \nonumber
\end{eqnarray}
it is possible to say that the generators $J_+,J_-,J_z$ in (6)
correspond to the Cartesian representation of the $SU(2)$ algebra.
Following \cite{19}, on can introduce the representation in
spherical coordinates via the polar decomposition of (6) of the
form
\begin{eqnarray}
J_+=J_r \epsilon , \quad \quad J_r=J_r^+, \quad \quad \epsilon
\epsilon^+ ={\bf 1}, \label{7}
\end{eqnarray}
where the Hermitian operator $J_r$ corresponds to the radial
contribution, white $\epsilon$ gives the exponential of the
azimuthal phase operator. It is a straightforward matter to show
that $\epsilon$ can be represented by the following $(2j+1) \times
(2j+1)$ matrix
\begin{eqnarray}
\epsilon = \left( \begin{array}{cccccc} 0 & 1 & 0 & 0 & \cdots & 0
\\ 0 & 0 & 1 & 0 & \cdots & 0 \\ \vdots & \vdots & \vdots & \vdots
& \vdots & \vdots \\ 0 & 0 & 0 & 0 & \cdots & 1 \\ e^{i \psi} & 0
& 0 & 0 & \cdots & 0 \end{array} \right) \label{8}
\end{eqnarray}
in the $2j+1$-dimensional Hilbert space. Here $\psi$ is an
arbitrary real parameter (reference phase). The eigenstates of the
operator (8)
\begin{eqnarray}
\epsilon | \phi_n^{(j)} \rangle =e^{i \phi_n^{(j)}} | \phi^{(j)}_n
\rangle, \quad n=1, \cdots , (2j+1), \label{9}
\end{eqnarray}
form the basis of the so-called phase states
\begin{eqnarray}
| \psi_n^{(j)} \rangle = \frac{1}{\sqrt{2j+1}} \sum_{k=0}^{2j}
e^{ik \phi_n^{(j)}}| \psi_k \rangle \label{10}
\end{eqnarray}
dual with respect to the basis of individual states $| \psi_k
\rangle$ of the Hilbert space.

As a physical example of some considerable interest, consider now
the system of the two identical two-level atom interacting with
the single cavity photon (see \cite{18}). If the cavity photon is
absorbed by either atom, the atomic subsystem  can be observed in
the following  states
\begin{eqnarray}
| \psi_1 \rangle = |e_1g_2 \rangle , \quad | \psi_2 \rangle =
|g_1e_2 \rangle , \label{11}
\end{eqnarray}
where $|e_1g_2 \rangle = |e_1 \rangle \otimes |g_2 \rangle$ and
$|e \rangle$ and $|g \rangle$ denote the excited and ground atomic
states, respectively. The subscript marks the atom. Using the
atomic basis (11), we can construct the following representation
of the $SU(2)$ algebra:
\begin{eqnarray}
J_+= |e_1g_2 \rangle \langle g_1e_2 |, \quad J_-=|g_1e_2 \rangle
\langle e_1g_2|, \nonumber \\ J_3= \frac{1}{2} (|e_1g_2 \rangle
\langle e_1g_2|- |g_1e_2 \rangle \langle g_1e_2 |). \label{12}
\end{eqnarray}
This representation formally corresponds to (6) at the spin
$j=1/2$. Then, the corresponding exponential of the phase operator
(8) takes the form
\begin{eqnarray}
\epsilon =|e_1g_2 \rangle \langle g_1e_2|+e^{i \psi}|g_1e_2
\rangle \langle e_1g_2|. \label{13}
\end{eqnarray}
In turn, the phase states (9) and (10) are
\begin{eqnarray}
| \phi_{\pm} \rangle = \frac{1}{\sqrt{2}} (|e_1g_2 \rangle +e^{i
\phi_{\pm}}|g_1e_2 \rangle ), \label{14} \\ \phi_{\pm}= \psi /2
+(1 \mp 1) \pi /2 . \nonumber
\end{eqnarray}
It is easily seen that the phase states (14) form the set of
entangled atomic states in the two-atom system under
consideration. Definitely, these states obey the nonseparability
condition. It is also seen that (14) coincides with the
maximally entangled states (2) of Ref. \cite{18} when the
reference phase $\psi =0$.

Consider now a general $2n+n$ system at $n \geq 1$. Then, the
maximum excited atomic states
\begin{eqnarray}
| \psi_i \rangle = | \{ e \}_n, \{ g \}_n \rangle,  \label{15}
\end{eqnarray}
can be used to construct a representation of the $SU(2)$ algebra
(6) of spin $j$ defined in (1). Here $i=1,2, \cdots ,N$ and
\begin{eqnarray}
N=2j+1= \left( \begin{array}{c} 2n \\ n \end{array} \right)
\nonumber
\end{eqnarray}
is the total number of such a states. In the basis (15), we can
construct the polar decomposition of the $SU(2)$ algebra of spin
(1) and the corresponding exponential of the phase operator (8)
and the phase states (10). Let us rename the states (15) as
follows
\begin{eqnarray}
| \psi_k \rangle \rightarrow | \psi_{k'} \rangle , \quad k' \equiv
k-1=0, \cdots ,N-1. \nonumber
\end{eqnarray}
Then, the $SU(2)$ phase states (10) take the form
\begin{eqnarray}
| \phi_k \rangle= \frac{1}{\sqrt{N}} \sum_{k'=0}^{N-1} e^{ik'
\phi_k}| \psi_{k'} \rangle , \label{16}
\end{eqnarray}
where
\begin{eqnarray}
\phi_k = ( \psi +2k \pi )/N. \nonumber
\end{eqnarray}
These states (16) form a basis dual with respect to (15) and
spanning the Hilbert space of the maximum excited atomic states in
the $2n+n$ system under consideration. By construction, the phase
states (16) are nonseparable with respect to contributions of
individual atoms and thus entangled \cite{24}. Let us stress that
the choice of the phase factors in (16) is irrelevant to
entanglement, which holds for arbitrary phase factors. This choice
is caused by the aspiration for getting the dual with respect to
(15) basis of entangled states.

It is easily seen that the states (16) obey the condition (4). In
fact, the action of  the  flip-operators $\sigma_{1,2}^{(\ell)}$
in (3) on the states (16) leads to the change of the number of
either excited or de-excited atoms:
\begin{eqnarray}
\sigma_{1,2}^{( \ell )}| \psi_k \rangle \rightarrow \left\{
\begin{array}{cc} | \{ e \}_{n-1} , \{ g \}_{n+1} \rangle & \ell \in \{ g \} \\ | \{ e
\}_{n+1} , \{ g \}_{n-1} \rangle & \ell \in \{ e \} \end{array}
\right. \nonumber
\end{eqnarray}
and therefore $\langle \sigma_{1,2}^{(\ell)} \rangle =0$ in the
case of averaging over the states (16). Since each state (15)
contains equal number of excited and de-excited atoms, the action
of the parity operator in (3) on the phase states (16) should lead
to the state which differ from (16) by the multiplication of a
certain $n$ terms by the factor of $-1$. Hence
\begin{eqnarray}
\langle \sigma_3^{(\ell)} \rangle = \frac{1}{N} \left(
\sum_{i=1}^{N/2} 1- \sum_{i=N/2+1}^N 1 \right) =0. \nonumber
\end{eqnarray}
By construction, $N$ is always an even number.

Thus, the $SU(2)$ phase states (16), corresponding to the maximum
excited atomic states in the $2n+n$ system, are entangled because
they are nonseparable and, at the same time, obey the condition
(4) for the local measurements. In the next Section, we show that
the states (16) manifest violation of classical realism as well.

Before we begin to discuss this subject, let us note that the
$SU(2)$ phase states of the atomic system under consideration with
integer spin do not provide the entanglement. Consider as an
example the system of three identical two-level atoms, interacting
with a single cavity photon. There are the three excited atomic
states
\begin{eqnarray}
|e_1g_2g_3 \rangle, \quad |g_1e_2g_3 \rangle, \quad |g_1g_2e_3
\rangle \label{17}
\end{eqnarray}
and the three dual phase states of the type of (16)
\begin{eqnarray}
| \psi_k \rangle = \frac{1}{\sqrt{3}} (|e_1g_2g_3 \rangle +e^{i
\phi_k}|g_1e_2g_3 \rangle +e^{2i \phi_k}|g_1g_2e_3 \rangle ).
\label{18}
\end{eqnarray}
It is clear that the states (18) are the phase states of spin
$j=1$. Here
\begin{eqnarray}
 \phi_k =( \psi +2k \pi)/3, \quad k=0,1,2. \nonumber
\end{eqnarray}
It is easily seen that the phase states (18) cannot be factorized
with respect to atoms. At the same time, the average of the parity
operator $\sigma_3^{(\ell)}$ in (3) over the states (18) is
\begin{eqnarray}
\forall k, \ell \quad \langle \psi_k | \sigma_3^{(\ell)} | \psi_k
\rangle =- \frac{1}{3}, \nonumber
\end{eqnarray}
although the averages of the flip-operators are
\begin{eqnarray}
\forall k, \ell \quad \langle \psi_k | \sigma_{1,2}^{(\ell)} |
\psi_k \rangle =0. \nonumber
\end{eqnarray}
Thus, the nonseparable states (18) do not obey the condition (4).
At the same time, these states do not manifest the maximum
entanglement as well. Let us stress that the nonseparability is
not a sufficient condition of maximum entanglement \cite{24}. For
example, from the measurement of the state of the first atom we
can only learn that either the atoms $2$ and $3$ are both in the
ground state with reliability or they are in the two-atom
entangled state of the type discussed in \cite{18}. Similar result
can be obtained for the system of three atoms, interacting with
two cavity photons. The only maximum entangled state of the system
of three atoms is provided by the superposition of GHZ states (5).

\section{The $4+2$-system}

To show that the phase states (16) of a $2n+n$ system violate the
classical realism, consider the system of four identical two-level
atoms interacting with two cavity photons. The maximum excited
atomic states at $n=2$ are
\begin{eqnarray}
|e_1e_2g_3g_4 \rangle , \quad |e_1g_2e_3g_4 \rangle , \quad
|e_1g_2g_3e_4 \rangle , \nonumber \\  |g_1e_2e_3g_4 \rangle ,
\quad |g_1e_2g_3e_4 \rangle , \quad |g_1g_2e_3e_4 \rangle .
\label{19}
\end{eqnarray}
These orthonormal states form the six-dimensional basis of the
Hilbert space in which the representation of the generators (6)
has the form
\begin{eqnarray}
J_+= \sqrt{5} |e_1e_2g_3g_4 \rangle \langle e_1g_2e_3g_4|+
\sqrt{8} |e_1g_2e_3g_4 \rangle \langle e_1g_2g_3e_4| \nonumber \\+
3|e_1g_2g_3e_4 \rangle \langle g_1e_2e_3g_4|+ \sqrt{8}
|g_1e_2e_3g_4 \rangle \langle g_1e_2g_3e_4| \nonumber \\ +
\sqrt{5} |g_1e_2g_3e_4 \rangle \langle g_1g_2e_3e_4| , \nonumber
\\ J_3= \frac{5}{2} |e_1e_2g_3g_4 \rangle \langle e_1e_2g_3g_4|+
\frac{3}{2} |e_1g_2e_3g_4 \rangle \langle e_1g_2e_3g_4| \nonumber
\\ + \frac{1}{2} |e_1g_2g_3e_4 \rangle \langle e_1g_2g_3e_4| -
\frac{1}{2} |g_1e_2e_3g_4 \rangle \langle g_1e_2e_3g_4| \nonumber
\\ - \frac{3}{2} |g_1e_2g_3e_4 \rangle \langle g_1e_2g_3e_4|-
\frac{5}{2} |g_1g_2e_3e_4 \rangle \langle g_1g_2e_3e_4| \nonumber
\end{eqnarray}
By construction, they describe the spin $j=5/2$ system. In turn,
the exponential of the phase operator (8) takes the form
\begin{eqnarray}
\epsilon =|e_1e_2g_3g_4 \rangle \langle
e_1g_2e_3g_4|+|e_1g_2e_3g_4 \rangle \langle e_1g_2g_3e_4|
\nonumber \\ +|e_1g_2g_3e_4 \rangle \langle
g_1e_2e_3g_4|+|g_1e_2e_3g_4 \rangle \langle g_1e_2g_3e_4|
\nonumber \\ +|g_1e_2g_3e_4 \rangle \langle g_1g_2e_3e_4|+e^{i
\psi}|g_1g_2e_3e_4 \rangle \langle e_1e_2g_3g_4|. \nonumber
\end{eqnarray}
Then, the six phase states (9) have the form (16) with $N=6$ and
\begin{eqnarray}
\phi_k = \frac{\psi}{6} + \frac{k \pi}{3} , \quad k=0,1, \cdots
,5. \label{20}
\end{eqnarray}
As well as (16), these states are nonseparable and hence entangled
and obey the condition (4) for local variables.

To show that these phase states violate the classical realism, let
us first represent the states (16) at $N=6$ in the following way
\begin{eqnarray}
| \phi_k \rangle = \frac{1}{\sqrt{3}} (| \chi_{1k} \rangle +e^{i
\phi_k}| \chi_{2k} \rangle +e^{2i \phi_k}| \chi_{3k} \rangle ),
\label{21}
\end{eqnarray}
where
\begin{eqnarray}
| \chi_{1k} \rangle & =  & \frac{1}{\sqrt{2}} (|e_1e_2g_3g_4
\rangle +e^{5i \phi_k}|g_1g_2e_3e_4 \rangle ), \nonumber \\ |
\chi_{2k} \rangle & = & \frac{1}{\sqrt{2}} (|g_1e_2e_3g_4 \rangle
+ e^{3i \phi_k}|e_1g_2g_3e_4 \rangle ) , \nonumber \\ | \chi_{3k}
\rangle & = & \frac{1}{\sqrt{2}} (|g_1e_2g_3e_4 \rangle + e^{i
\phi_k}|e_1g_2e_3g_4 \rangle ). \label{22}
\end{eqnarray}
It is easily seen that each set of six states $| \chi_{pk}
\rangle$ with $p=1,2,3$ and $k=0, \cdots ,5$ consists of the
nonseparable and hence entangled states. Consider, for example,
the states $| \chi_{1k} \rangle$ in (22). Because of the
definition of the phase angle $\phi_k$ at $N=6$, they consist of
the three sets of the pairwise orthogonal states
\begin{eqnarray}
\{ | \chi_{10} \rangle , | \chi_{13} \rangle \} , \quad \{ |
\chi_{11} \rangle , | \chi_{14} \rangle \} , \quad \{ | \chi_{12}
\rangle , | \chi_{15} \rangle \} . \nonumber
\end{eqnarray}
It is also seen that the second and third sets here are obtained
from the first set by the successive rotations of the reference
frame.

Now the violation of classical realism can be proved through the
use of the GHZ theorem \cite{10}. Consider first the state $|
\chi_{10} \rangle$ in (22). It is easy to verify that this state
obey the conditions
\begin{eqnarray}
\forall i, \ell \quad \quad \bigotimes_{\ell =1}^4
\sigma_{i}^{(\ell)} | \chi_{10} \rangle = | \chi_{10} \rangle
\label{23}
\end{eqnarray}
and
\begin{eqnarray}
\sigma_1^{(1)} \sigma_{1}^{(2)} \sigma_2^{(3)} \sigma_2^{(4)} |
\chi_{10} \rangle & = &  -| \chi_{10} \rangle , \nonumber \\
\sigma_2^{(1)} \sigma_2^{(2)} \sigma_1^{(3)} \sigma_1^{(4)} |
\chi_{10} \rangle & = & -| \chi_{10} \rangle , \nonumber \\
\sigma_1^{(1)} \sigma_2^{(2)} \sigma_1^{(3)} \sigma_2^{(4)} |
\chi_{10} \rangle & = & | \chi_{10} \rangle , \nonumber \\
\sigma_1^{(1)} \sigma_2^{(2)} \sigma_2^{(3)} \sigma_1^{(4)} |
\chi_{10} \rangle & = & | \chi_{10} \rangle , \nonumber \\
\sigma_2^{(1)} \sigma_1^{(2)} \sigma_2^{(3)} \sigma_1^{(4)} |
\chi_{10} \rangle & = & | \chi_{10} \rangle , \nonumber \\
\sigma_2^{(1)} \sigma_1^{(2)} \sigma_1^{(3)} \sigma_2^{(4)} |
\chi_{10} \rangle & = & | \chi_{10} \rangle . \label{24}
\end{eqnarray}
It is possible to say that these equalities (23) and (24) express
a kind of EPR "action at distance" in the maximum excited states
of the system of four atoms interacting with two photons. In other
words, the correlations represented by (23) and (24) permit us
determine in a unique way the state of the fourth atom via
measurement of the states of other three atoms.

The operator equalities (23) and (24) can be used to obtain the
relations similar to those in the GHZ theorem. Following
\cite{10}, we have to assign the classical quantities
$m^{(\ell)}_i$ to the local operators. Here
\begin{eqnarray}
m^{(\ell)}_1,m^{(\ell)}_2= \pm 1. \nonumber
\end{eqnarray}
Then, it follows from (23) that
\begin{eqnarray}
\prod_{\ell =1}^4 m_1^{( \ell )}=1. \label{25}
\end{eqnarray}
At the same time, it follows from (24) that
\begin{eqnarray}
[ \sigma_1^{(1)} \sigma_1^{(2)} \sigma_2^{(3)} \sigma_2^{(4)}][
\sigma_{1}^{(1)} \sigma_2^{(2)} \sigma_1^{(3)} \sigma_{2}^{(4)} ][
\sigma_1^{(1)} \sigma_2^{(2)} \sigma_2^{(3)} \sigma_1^{(4)}]|
\chi_{10} \rangle \nonumber \\ =-| \chi_{10} \rangle . \nonumber
\end{eqnarray}
Employing the classical variables instead of the local operators
allows this to be cast into the form
\begin{eqnarray}
(m_1^{(1)})^3m_1^{(2)}(m_2^{(2)})^2m_1^{(3)}(m_2^{(3)})^2m_1^{(4)}(m_2^{(4)})^2=-1.
\nonumber
\end{eqnarray}
Since $(m_1^{( \ell)})^2=(m_2^{( \ell)})^2 =1$, we get an
equivalent equality
\begin{eqnarray}
m_1^{(1)}m_1^{(2)}m^{(3)}_1m^{(4)}_1 = -1, \nonumber
\end{eqnarray}
which contradicts (25). Hence, the state $| \chi_{10} \rangle$ in
(22) obey the GHZ theorem. Similar result can be obtained for all
other states in (22) and hence, for the phase states (21).

Our consideration so far have applied to the local measurements
touching on a single atom. We now note that the phase states (21)
allow another kind of entanglement in the case of pairwise
measurement. Consider again the state $| \chi_{10} \rangle$ in
(22) and assume that the measurements $a$ and $b$ corresponds to a
pair of atoms:
\begin{eqnarray}
a & = & \cos \theta_a |e_1e_2 \rangle \langle e_1e_2|+ \sin
\theta_a (|e_1e_2 \rangle \langle g_1g_2| \nonumber
\\ & + & |g_1g_2 \rangle \langle e_1e_2 |)- \cos \theta_a |g_1g_2
\rangle \langle g_1g_2|, \nonumber \\ b & = & \cos \theta_b
|e_3e_4 \rangle \langle e_3e_4| + \sin \theta_b (|e_3e_4 \rangle
\langle g_3g_4| \nonumber \\ & + & |g_3g_4 \rangle \langle
e_3e_4|)- \cos \theta_b |g_3g_4 \rangle \langle g_3g_4|.
\label{26}
\end{eqnarray}
Assume now that we make the two measurements $a$ and $a'$ with the
angles $\theta_1 = \pi$ and ${\theta'}_a = \pi /2$ and the two
more measurements $b$ and $b'$ with the angles ${\theta'}_b =-
\theta_b$, respectively. Then, the averaging over the state $|
\chi_{10} \rangle$ gives
\begin{eqnarray}
\langle ab \rangle = \langle ab' \rangle = \cos \theta_b , \quad
\langle a'b \rangle= \sin \theta_b =- \langle a'b' \rangle .
\nonumber
\end{eqnarray}
Employing the CHSH inequality \cite{33}
\begin{eqnarray}
| \langle ab \rangle + \langle a'b \rangle + \langle a'b' \rangle
- \langle ab' \rangle | \leq 2 \label{27}
\end{eqnarray}
then gives
\begin{eqnarray}
| \cos \theta_b - \sin \theta_b | \leq 1. \nonumber
\end{eqnarray}
Violation  of this inequality and hence, of the classical realism
occurs at small negative $\theta_b$, when we can put
\begin{eqnarray}
| \cos \theta_b - \sin \theta_b | \sim 1+| \theta_b |>1. \nonumber
\end{eqnarray}
Similar consideration can be done for all over states in (22)
through the use of proper pairwise measurements. At the same time,
the phase states (21) do not manifest entanglement with respect to
the pairwise measurements.

The phase states (16) for the $6+3$, $8+4$, $\cdots $ systems,
corresponding to the spin (1) equal to $19/2,69/2, \cdots$,
respectively, can be considered as above.

\section{Initial conditions and atomic entanglement}

It is clear that the evolution of the $2n+n$ system strongly
depends on the choice of initial conditions. To trace the proper
choice leading to the atomic entanglement, let us ignore the
relaxation processes. Then, the steady-state evolution of the
$2n+n$ system under consideration is governed by the Hamiltonian
\begin{eqnarray}
H= \Delta a^+a + \omega_0 {\cal N}+ \gamma \sum_{ \ell }
(R^+_{\ell} a+a^+R_{\ell} ). \label{28}
\end{eqnarray}
Here $\Delta$ is the cavity detuning, $ \omega_0$ is the atomic
transition frequency, $\gamma$ is the atom-field coupling
constant, operators $a$ and $a^+$ describe the cavity photons,
\begin{eqnarray}
{\cal N}=a^+a+ \sum_{\ell} |e_{\ell} \rangle \langle e_{\ell}|
\bigotimes_{\ell' \neq \ell} {\bf 1}^{(\ell)}, \nonumber
\end{eqnarray}
and the atomic operators are defined as follows
\begin{eqnarray}
R^+_{\ell}=|e_{\ell} \rangle \langle g_{\ell}| \bigotimes_{\ell'
\neq \ell} {\bf 1}^{(\ell')}. \nonumber
\end{eqnarray}
Here ${\bf 1}^{(\ell)}$ denotes the unit operator in the
two-dimensional Hilbert space of the $\ell^{th}$ atom. It is seen
that $[{\cal N},H]=0$. It is also seen that the atomic operators
are similar, in a certain sense, to the local operators (3). In
fact
\begin{eqnarray}
R_{\ell}^{\pm}= \frac{ \sigma_1^{(\ell)} \pm i
\sigma_2^{(\ell)}}{2} . \nonumber
\end{eqnarray}

Consider first the case of two atoms and single cavity photon when
$\ell =1,2$ and the Hamiltonian (28) coincides with that of Ref.
\cite{18}. For simplicity, we use here the same coupling constant
$\gamma$ for both atoms. Our consideration can easily be
generalized on the case of coupling constant depending on the
atomic position. Let us note that, in the case of only two atoms,
the Hamiltonian (28) can be represented as follows
\begin{eqnarray}
H \rightarrow H_{\phi} = \Delta a^+a+ \omega_0 {\cal N}_{\phi}+
\gamma \sqrt{2} ({\cal R}^+a+a^+{\cal R}), \label{29}
\end{eqnarray}
where
\begin{eqnarray}
{\cal N}_{\phi} = a^+a+ \sum_{k = \pm 1} | \phi_k \rangle \langle
\phi_k | \nonumber
\end{eqnarray}
and
\begin{eqnarray}
{\cal R}^+=| \phi_+ \rangle \langle g_1g_2|. \nonumber
\end{eqnarray}
Here $| \phi_{\pm} \rangle$ denote the phase states (14).

Using the Hamiltonian (29) as the generator of evolution, for the
time dependent wave function we get
\begin{eqnarray}
| \Psi (t) \rangle =e^{-iH_{\phi}t}| \Psi (0) \rangle =[ C_-(t)|
\phi_- \rangle \nonumber \\  +C_+(t) | \phi_+ \rangle ] \otimes |0
\rangle_{ph}  +C(t)|g_1g_2 \rangle \otimes |1 \rangle_{ph} ,
\label{30}
\end{eqnarray}
where $| \cdots \rangle_{ph}$ denotes the states of the cavity
field. The coefficients $C_{\pm}(t)$ and $C(t)$ in (30) are
completely determined by the initial conditions and normalization
condition.

It is easily seen that the state $| \phi_- \rangle \otimes |0
\rangle_{ph}$ is the eigenstate of the Hamiltonian (29). Hence, at
\begin{eqnarray}
C_-(0)=1, \quad \quad C_+(0)=C(0)=0, \nonumber
\end{eqnarray}
the atomic phase state $| \phi_- \rangle$ in (14) provides the
stationary, maximum entangled atomic state in the system under
consideration \cite{18}. At the same time, it is not very clear
how to prepare such a state.

Therefore we consider a more realistic initial state provided by
excitation of either atom, while the cavity field is in the vacuum
state. To realize such a state, we can assume, for example, that
one of the atoms (initially de-excited) is trapped in the cavity,
while the second atom (initially excited) slowly passes through
the cavity like in the experiments discussed in \cite{14,15}.
assume for definiteness that
\begin{eqnarray}
| \Psi (0) \rangle = |e_1g_2 \rangle \otimes |0 \rangle_{ph} .
\label{31}
\end{eqnarray}
Then, the coefficients of the wave function (30) take the form
\begin{eqnarray}
C_-(t) & = & \frac{1}{\sqrt{2}} e^{-i \omega_0 t} , \nonumber \\
C_+(t) & = & \frac{1}{\sqrt{2}} \left( \cos \Omega t + \frac{i
\Delta}{2 \Omega} \sin \Omega t \right) e^{-i( \omega_0 + \Delta
/2)t}, \nonumber \\ C(t) & = & - \frac{i \gamma}{\Omega} e^{-i(
\omega_0 + \Delta /2)t} \sin \Omega t , \nonumber
\end{eqnarray}
where $\Omega =[2 \gamma^2 +( \Delta /2)^2]^{1/2}$. At first site,
the probabilities
\begin{eqnarray}
P_{\pm}(t)=| \langle 0|_{ph} \otimes \langle \phi_{pm} | \Psi (t)
\rangle |^2 = |C_{\pm}(t)|^2 \nonumber
\end{eqnarray}
to observe the states (14) corresponding to the maximum atomic
entanglement, are
\begin{eqnarray}
P_-(t) & = & \frac{1}{2} , \nonumber \\ P_+(t) & = &
\frac{\Delta^2}{8 \Omega^2} + \frac{\gamma^2}{\Omega^2} \cos^2
\Omega t \leq \frac{1}{2} , \nonumber
\end{eqnarray}
respectively. At the same time, the absence of photon counts,
which is considered in \cite{18} as a sign of the atomic
entanglement, corresponds here to the case when both probabilities
$P_{\pm}(t_k)=1/2$ at a certain time $t_k$. In other words, the
mutually orthogonal entangled states (14) have the same
probability to be observed at $t=t_k$. This means that there is no
atomic entanglement at all but we definitely know which atom is in
the excited state.

Consider one more realistic initial state when both atoms are
trapped in the cavity in de-excited state, while the cavity field
contains a photon:
\begin{eqnarray}
| \Psi (0) \rangle = |g_1g_2 \rangle \otimes |1 \rangle_{ph} .
\label{32}
\end{eqnarray}
Then, for all times we get $C_-(t)=0$ and
\begin{eqnarray}
C_+(t) & = & - \frac{i \gamma \sqrt{2}}{\Omega} e^{-i( \omega_0 +
\Delta /2)t} \sin \Omega t, \nonumber \\ C(t) & = & \left( \cos
\Omega t - \frac{i \Delta}{2 \Omega} \sin \Omega t \right) e^{-i(
\omega_0 + \Delta /2)t} . \nonumber
\end{eqnarray}
Hence, under this initial condition, the entangled state $| \phi_-
\rangle$ cannot be achieved at all, while the second entangled
state $| \phi_+ \rangle$ in (14) can be achieved. It is seen that,
in the case of initial state (32), the probability to detect the
photon is
\begin{eqnarray}
P_{ph}(t)=|C(t)|^2= \cos^2 \Omega t + \frac{\Delta^2}{4 \Omega^2}
\sin^2 \Omega t. \nonumber
\end{eqnarray}
This expression takes the minimum value
\begin{eqnarray}
 \min P_{ph}=P_{ph}(t_m)= \frac{\Delta^2}{4 \Omega^2} \nonumber
\end{eqnarray}
at $t=t_m= \pi(2m+1)/ 2\Omega$, $m=0,1, \cdots$. At the same time
$t_m$, the probability to have the entangled atomic state $|
\phi_+ \rangle$ takes the maximum value
\begin{eqnarray}
P_+(t_m)=|C_+(t_m)|^2= \frac{2 \gamma^2}{2 \gamma^2 +( \Delta
/2)^2}. \nonumber
\end{eqnarray}
It is seen that the pure atomic entanglement with $P_+(t_m)=1$ is
realized at $t=t_m$ only in the absence of the cavity detuning
when $\Delta \rightarrow 0$.

The parasitic influence of the cavity detuning can be compensated
through the use of Kerr medium filling the cavity. In this case,
the Hamiltonian (28) should be supplemented by the term
\begin{eqnarray}
H_{\kappa}= \kappa (a^+a)^2, \nonumber
\end{eqnarray}
which leads to the following renormalization of the Rabi frequency
\begin{eqnarray}
\Omega \rightarrow \Omega_{\kappa}= \sqrt{2 \gamma^2 +( \Delta +
\kappa )^2 /4}. \nonumber
\end{eqnarray}
Then, the proper choice of the Kerr parameter $\kappa = -\Delta$
should lead to the pure entangled atomic state $| \phi_+ \rangle$
at a certain times.

Consider now the case of four atoms and two photons. In contrast
to the previous case, neither phase state in (21) is an eigenstate
of the Hamiltonian (28). Then, the choice of the initial state
either as a state with two excited atoms or as a state with one
excited atom plus cavity photon does not lead to a pure atomic
entanglement. As in the case of two atoms, the pure atomic
entanglement can be reached under the choice of the state with the
absence of the atomic excitations in the initial state. The
influence of the cavity detuning can be compensated by the
presence of Kerr medium as well as in the case of two atoms.

\section{Conclusion}

Let us briefly discuss the obtained results. For the system of two
identical two-level atoms interacting with a single photon has
been proposed in \cite{18} it is shown that the maximum entangled
atomic states are represented by the $SU(2)$ phase states of spin
$1/2$. Moreover, it is shown that the $SU(2)$ phase states of the
half-integer spin $j$ (1) form a certain class of maximum
entangled atomic states in the system of $2n$ atoms interacting
with $n$ photons. In particular, the violation of classical
realism is shown.

It should be noted in this connection that the above considered
$SU(2)$ phase states do not represent a unique way to construct
the maximum entangled states in the multi-atom systems and that
some other symmetries, for example the $SU(\cal N)$ can
be considered as well. Moreover, in some cases the $SU(2)$ phase
states cannot be used to determine the maximum entangled states at
all. Consider for example the case of two identical two-level
atoms interacting with two photons, when the atomic subsystem can
be specified by the four states
\begin{eqnarray}
|e_1e_2 \rangle , \quad |e_1g_2 \rangle , \quad |g_1e_2 \rangle ,
\quad  |g_1g_2 \rangle . \nonumber
\end{eqnarray}
By performing a similar analysis to that described in section II,
it is easy to construct the corresponding set of the $SU(2)$ phase
states
\begin{eqnarray}
| \phi_k \rangle = \frac{1}{2} (|e_1e_2 \rangle +e^{i
\phi_k}|e_1g_2 \rangle +e^{2i \phi_k}|g_1g_2 \rangle +e^{3i
\phi_k}|g_1e_2 \rangle ), \nonumber \\ \phi_k = \frac{\psi}{4} +
\frac{k \pi}{2} , \quad k=0,1,2,3, \nonumber
\end{eqnarray}
which do not manifest the maximum entanglement. At the same time,
the general criterion (4) permits us to determine infinitely many
maximum entangled states in this case \cite{23}. An example is
provided by the following set of orthonormal maximum entangled
states
\begin{eqnarray}
| \psi_1 \rangle  =  \frac{1}{2} (|e_1e_2 \rangle +|g_1g_2 \rangle
+ i|e_1g_2 \rangle +i|g_1e_2 \rangle ), \nonumber \\ |\psi_2
\rangle = \frac{1}{2} (|e_1e_2 \rangle -|g_1g_2 \rangle -i|e_1g_2
\rangle +i|g_1e_2 \rangle ), \nonumber \\ | \psi_3 \rangle =
\frac{1}{2} (i|e_1e_2 \rangle +i|g_1g_2 \rangle +|e_1g_2 \rangle
+|g_1e_2 \rangle ), \nonumber
\\ | \psi_4 \rangle = \frac{1}{2} (-i|e_1e_2 \rangle +i|g_1g_2 \rangle
+|e_1g_2 \rangle -|g_1e_2 \rangle ). \nonumber
\end{eqnarray}
In fact, the Eq. (4) gives a general condition \cite{23}, while
the $SU(2)$ phase states can manifest the maximum entanglement
only under a certain  condition (special choice of the effective
spin (1)).

Nevertheless, the $SU(2)$ phase states considered in sections II
and III represent an important example of the atomic entangled
states. First of all, they can be easily realized in the atomic
systems in a cavity. In fact, these states have a simple physical
meaning. In addition to (9), the $SU(2)$ phase states can be
defined to be the eigenstates of the Hermitian "cosine" operator
\cite{19,22}
\begin{eqnarray}
C = \frac{1}{2} (\epsilon + \epsilon^+ ) ,  \nonumber
\end{eqnarray}
where $\epsilon$ is defined by Eq. (8). This operator $C$ can be
considered as a "Hamiltonian", describing the correlations between
the different atoms. For example, in the case of the two atoms
interacting with the single photon, the operator $C$ takes the
form
\begin{eqnarray}
C= \sigma_+^{(1)} \sigma_-^{(2)}+ \sigma_-^{(2)} \sigma_+^{(1)},
\label{33}
\end{eqnarray}
where
\begin{eqnarray}
\sigma_{\pm}^{(\ell)} = \frac{\sigma_1^{(\ell)} \pm i
\sigma_2^{(\ell)}}{2} . \nonumber
\end{eqnarray}
The operator structure of (33) coincides with that of the
so-called model of plane rotator, which is a particular case of
the Heisenberg model of ferromagnetism widely used in statistical
physics \cite{35} and in quantum information theory \cite{36}.

Let us also stress that the $SU(2)$ phase states similar to those
considered in sections II and III, have been discussed recently in
the context of quantum coding \cite{37}.

It is also known that  the $SU(2)$ phase states have direct
connection with the quantum description of polarization of
spherical photons emitted by the multipole transitions in atoms
and molecules \cite{21,22,38}. Therefore, the polarization
entanglement of photons can be examined in direct analogy to the
above discussed atomic entanglement \cite{39}. At the same time,
the consideration of spherical photons requires the use of more
quantum degrees of freedom.  Consider as an example the cascade
decay of a two-level atom specified by the transition \cite{40}
\begin{eqnarray}
|J=2,m=0 \rangle \rightarrow |J'=0,m'=0 \rangle . \nonumber
\end{eqnarray}
Here $J,J'$ and $m,m'$ denote the angular momentum and projection
of the angular momentum of the excited and ground atomic states,
respectively. This transition gives rise to an entangled photon
twins \cite{40}. Each photon carries spin $1$, but because of the
conservation of the angular momentum in the process of radiation,
the sum of projections of the angular momenta of the two photons
should be equal to zero. Denoting the state of a photon with given
$m$ by $|m \rangle$, we get the three possible states of the
photon subsystem
\begin{eqnarray}
|+1 \rangle \otimes |-1 \rangle , \quad |0 \rangle \otimes |0
\rangle , \quad |-1 \rangle \otimes |+1 \rangle . \nonumber
\end{eqnarray}
These three "individual" states can be used to construct the dual
basis of the $SU(2)$ phase states \cite{21}
\begin{eqnarray}
| \phi_k \rangle = \frac{1}{\sqrt{3}} (|+1 \rangle \otimes |-1
\rangle +e^{i
\phi_k}|0 \rangle \otimes |0 \rangle \nonumber \\ +e^{2i\phi_k}|-1 \rangle
\otimes |+1 \rangle , \nonumber \\
\phi_k= \frac{\psi +2k \pi}{3}, \quad k=0,1,2  \label{34}
\end{eqnarray}
similar to (18). It can be easily seen that these states manifest
the maximum entanglement.

Similar entangled states have been discussed in the context of the
so-called biphoton excitations \cite{41} (photon pairs in
symmetric Fock states). They can also be used in quantum
cryptography \cite{42}.

Let us stress that the general condition of the type of (4) is
also valid in the case of states (34). However, the definition of
local measurement should be changed in this case. Because of the
number of degrees of freedom per photon is equal to three, the
Hermitian operators associated with the $SU(3)$ group should be
considered instead of the infinitesimal generators of the $SL(2)$
group. For example, the set of the Stokes operators of the Ref.
\cite{21}, corresponding to the representation of the $SU(3)$
subalgebra in the Weyl-Heisenberg algebra of spherical photons,
can be used to define the complete set of local measurements in
this case.

It is shown in section IV that the realization of a pure atomic
entanglement in the $2n+n$-type atom + photon systems strongly
depends on the choice of initial state. Viz, the entangled states
can be reached in the process of steady-state evolution only if
all $2n$ atoms are initially in the de-excited states, while the
cavity contains just $n$ photons. This condition has an
intuitively clear explanation: the excitations of different atoms
have the same probability and therefore each photon in the
$2n+n$-system is shared with a couple of atoms.

It is also shown in section IV that the presence of the cavity
detuning hampers the creation of a pure entangled atomic state.
This negative effect can be compensated through the use of Kerr
medium in the cavity.

We now note that the practical realization of a long-lived,
maximum entanglement in a quantum mechanical system strongly
depends on the interaction between this system and environment.
The point is that the state of a closed quantum mechanical system
changes periodically, providing the maximum entanglement as an
instant event only at a certain times (see section IV). Such a
periodicity is caused by a finite number of degrees of freedom in
the system. To destruct such a periodicity, it is necessary to
connect the system to a "heat bath", which would tune in the
system to a required state. In Ref. \cite{18}, it has been
proposed to support the atomic entanglement by the cavity losses.
In this case, the absence of the photon counting outside the
cavity can be associated with the existence of the entangled
atomic state in the cavity.

Let us stress that an advantage of the use of the $SU(2)$ phase
states as the maximum entangled atomic states consists in the
simple preparation of the initial states discussed in section IV.

In view of realization of atomic entanglement with present
experimental technique, it seems to be more convenient if the
existence of entangled state in a cavity would manifest itself via
a signal photon rather than the absence of photon leakage from the
cavity. In this case, there should be at least the two modes such
that one of them (the cavity mode) provides the correlation
between the atoms, while the second can freely leave the resonator
to signalize the existence of the entanglement. Such a process can
be realized through the use of the Raman process in atoms shown in
Fig. 1 (e.g., see \cite{43}). Here the dipole transitions are
allowed between the levels $1$ and $2$ and $2$ and $3$, while
forbidden between $1$ and $3$ because of the parity conservation.
In the simplest case, we should assume that the two identical
atoms of this type are located in a cavity, which has a very high
quality with respect to the pumping mode $\omega_P$, while the
Stokes photons with frequency $\omega_{Sk}$ can leak away freely.

Assume that the atoms are initially in the ground state $1$, the
Stokes field is in the vacuum state, and the pump field consists
of a single photon. The evolution of the system can lead to the
absorption of the cavity photon by either atom with further
emission of the Stokes photon, which leaves the cavity. After
that, the atoms are in entangled state, corresponding to the
excitation of the atomic level $3$ shared between the atoms. Since
the inverse process cannot be realized without assistance of the
Stokes photon, such a state represents a durable atomic entangled
state.

It is clear that the above consideration of the atomic
entanglement in the multi-atom system can be generalized with easy
on the case of Raman process in atoms. In other words, the $SU(2)$
phase states similar to (16) form the class of the maximum
entangled atomic states in the case of Raman-type processes in the
three-level atoms as well. An evident advantage of the use of the
Raman process is the long-lived maximum entanglement in atomic
subsystem.

One of the authors (A.S.) would like to thank Dr. A. Beige, Prof.
P.L. Knight, Prof. A. Vourdas, and Prof. A. Zeilinger for useful
discussions.

\begin{figure}
\caption{Atomic Raman-type interaction with pump (P) and Stokes
(S) photons.}
\end{figure}

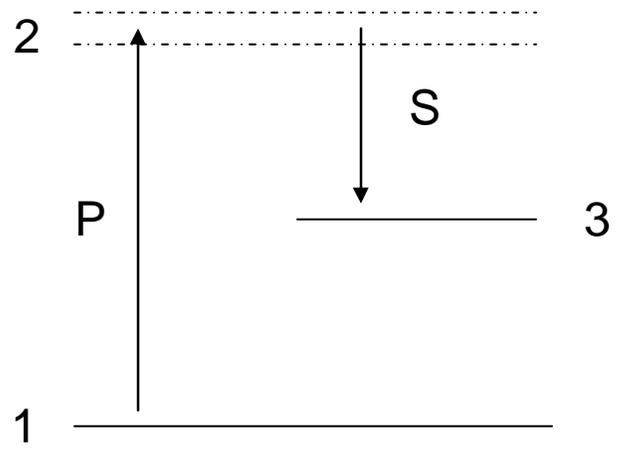
\end{document}